\documentclass[twocolumn,showpacs,preprintnumbers,amsmath,amssymb]{revtex4}

\usepackage{latexsym}
\usepackage{amssymb}
\usepackage{amsmath}

\def\longbar#1{#1\kern-0.7em\raise1.3ex\hbox{{$-$}}} 

\def\a{\alpha}  \def\g{\gamma} \def\d{\delta}
   \def\D{\Delta}
\def\m{\mu}

\def\pa{\partial}

\newcommand{\elR }[2]{{\frac{\delta^R   {#1}}{\delta   {#2}}}}
\newcommand{\elL }[2]{{\frac{\delta^L   {#1}}{\delta   {#2}}}}
\newcommand{\el  }[2]{{\frac{\delta     {#1}}{\delta   {#2}}}}

\def\Cb{\ensuremath{{\overline{\cC}}}}

\def\cA{{\cal A}} \def\cB{{\cal B}} \def\cC{{\cal C}} \def\cD{{\cal D}}

  \def\mR{{\mathbb{ R}}}

\def\half{\frac{1}{2}}

\usepackage{graphicx}
\usepackage{dcolumn}
\usepackage{bm}

\begin{document}

\title{The unphysical nature of the $SL(2,\mR)$ symmetry and
   its associated condensates in Yang-Mills theories}

\author{Mboyo Esole}
\author{Filipe Freire}
\affiliation{Instituut-Lorentz, Universiteit Leiden, P.\,O. Box
 9506, 2300 RA Leiden, The Netherlands} 

\begin{abstract}
BRST cohomology methods are used to explain the 
origin of the $SL(2,\mR)$ symmetry in Yang-Mills theories. 
Clear evidence is provided for the unphysical nature of this symmetry. 
This is obtained from the analysis of a local functional of mass
dimension two and constitutes a no-go statement 
for giving a physical meaning to condensates associated with the
symmetry breaking of $SL(2,\mR)$.
\end{abstract}


\maketitle
In recent years there has been a growing interest in
condensates of functionals of mass dimension two in $SU(N)$
Yang-Mills (YM) theories in four dimensions 
involving ghost fields\,\cite{Schaden:1999ew,Kondo:2001nq,Lemes:2002jv}.
The condensates are signaled by non-vanishing expectation values of
the functionals.
These studies have been done in the maximal Abelian (MA) gauge and the
generalized Lorentz or Curci-Ferrari (CF) gauge and
were motivated by the prospect that these condensates may
control the infrared divergences in perturbation theory and
consequently shed light on the origin of the mass gap for the gluon
excitations.

The interest in condensates involving ghosts goes back to the
suggestion that ghost-antighost pairs condense in the
MA gauge following the symmetry breaking
of a global $SL(2,\mR)$ symmetry of the gauge-fixed
action\,\cite{Schaden:1999ew}.
Also in the CF gauge the breaking of an $SL(2,\mR)$ symmetry has been
associated with ghost condensates\,\cite{Lemes:2002jv}.
The mass generated by this condensation scenario was later shown to be 
tachyonic\,\cite{Dudal:2002xe}. 
The origin and physical relevance of the links between the condensates,
the gauges used and the $SL(2,\mR)$ symmetry remained unclear.

It has then been suggested that in $SU(N)$ YM the mass dimension two
local functional 
\begin{equation}
\label{operator}
A_0 = \textstyle \int d^n x\,(\,\frac12 \cB_{\m }^A 
\cB^{A\m} +\a \cC^A{\Cb}{}^A), \,A=1\,\cdots\,N^2-1,
\end{equation}
might play an important role in the mass generation for gluons because
it is ``BRS-closed''\,\cite{Kondo:2001nq}.
In \eqref{operator} $\cB_{\m}^A$ is the gauge potential, $\cC^A$ and
$\Cb{}^A$ are,  
respectively, the ghost and antighost fields and $\alpha$ is 
a gauge-fixing parameter.
By local it is meant that the functional depends on the fields and a
finite number of their derivatives all of which are
evaluated at the same point in space-time.

In this letter we use  BRST cohomology methods in the
Batalin-Vilkovisky (BV) antifield 
formalism\,\cite{Batalin,HenneauxBook,Gomis,HenneauxReport} to explain the 
links between the functional $A_0$, the CF gauge and the $SL(2,\mR)$ symmetry.
It is important to realize that BRST
cohomology though it is a perturbative method can provide important 
information on non perturbative phenomena.  This is well illustrated
in the classification of anomalies \cite{HenneauxReport}.  Like anomalies, 
the formation of condensates is associated with symmetry breaking.
The difference here is that BRST cohomology is used to study the
physical relevance of the broken symmetry associated with the
condensate.
For convenience, we use the terminology
BRST only for the $s$-transformations in the antifield BV formalism
while BRS is used for the original Becchi-Rouet-Stora
transformation\,\cite{Becchi}.  

We present a clear picture of why no physical content 
should be attributed to condensates 
linked to the breaking of $SL(2,\mR)$, e.g.,
$\langle f^{ABC}\cC^B\cC^C\rangle$, $\langle f^{ABC}\Cb{}^B\Cb{}^C\rangle$ and 
$\langle f^{ABC}\cC^B\Cb{}^C\rangle$\,\cite{Lemes:2002jv}. 
This is obtained by studying the local functional $A_0$
given by \eqref{operator}. We note that $A_0$ is defined without
specifying any particular gauge, but it is known that it is \textit{only} 
in the CF gauge that $A_0$ is BRS and anti-BRS
closed modulo the equations of motion (EOM) of the gauge-fixed action.
However, this does not imply the existence of an observable
associated to $A_0$. Indeed, for local functionals the
on-shell BRS invariance $must$ be supplemented by appropriate
conditions that guarantee the isomorphism between the cohomology of
the on-shell BRS
symmetry and the BRST operator\,\cite{HenneauxGF1,HenneauxGF2}.
It follows from\,\cite{HenneauxGF2,Brandt} that the
non-existence of a local observable associated to an on-shell
(anti-)BRS closed local
functional is due to global symmetries of the
corresponding \textit{gauge-fixed action}. For $A_0$, these
symmetries are associated to the non-diagonal generators of
$SL(2,\mR)$. These
generators can not have any physical meaning in YM as they always
involve trivial elements of the cohomology\,\cite{HenneauxReport}.
Therefore, condensates associated to the symmetry breaking of
$SL(2,\mR)$ can not be linked to 
the mass generation in YM 
contrary to the suggestions
in\,\cite{Lemes:2002jv,Dudal:2003dp}.


Before starting our study of $A_0$ we summarize the salient
features of the BV formalism in the context of YM
theories\,\cite{Batalin,HenneauxBook,Gomis,HenneauxReport}.
The antifield formalism starts by enlarging the original space to
contain not only the gauge fields $\cB^A_\m$ and the ghosts
$\cC^A$, but also sources for their BRST
variations\,\cite{Zinn-Justin}  denoted respectively by $\cB^{*\m}_A$ and 
$\cC^*_A$. They are the {\em antifields} in the BV 
formalism and each of them has a 
Grassmann parity opposite to the corresponding field. 
An odd symplectic structure $(\,.\,,\,.\,)$ called the antibracket
is defined on the extended phase space, so that 
the fields $\Phi^i=\{\cB_{\mu}^A,\cC^A\}$  are canonically
conjugate to the antifields  $\Phi^*_i=\{\cB^{*\mu}_A,\cC^*_A\}$ in
the sense that $(\Phi^i,\Phi^*_j)\equiv\delta^i_{\,j}$.
The antibracket is extended to any functionals $A=\!\int\!d^nx\,a$
and $B=\!\int\!d^nx\,b$, as
$(A,B) =\!\int\!d^nx\,(\elR{a}{\Phi^i}\elL{b}{\Phi^*_i}
-\elR{a}{\Phi^*_i}\elL{b}{\Phi^i})$,
where $\frac{\d^{L,R}}{\d z}$,
denote respectively left and right Euler-Lagrange derivatives.
In the context of the gauge independent formulation no antighosts or
auxiliary fields are needed.
The classical action $S_0$ is extended to a local functional $S$,
which  includes terms involving ghosts and antifields and is a
proper solution of the master equation $(S,S)=0$. 
This equation contains all the information about the
action, the  infinitesimal gauge transformations
and their algebra. For YM the minimal proper solution is
\begin{equation} 
S =S_0+\textstyle \int d^n x\,(\cB^{*\mu}_A \cD^{AB}_\mu \cC^B + 
\half \cC^*_{A} {f^{ABC}}\cC^C \cC^B)\label{master}\,,
\end{equation}
where $S_0$ is the YM action, $f^{ABC}$ are the gauge group structure
constants and $\cD_\mu^{AB}=\delta^{AB}\partial_\mu - 
f^{ABC} \cB_\mu^C$.  

The BRST operator is canonically generated by $S$
through the antibracket in the sense that  $sA = (S,A)$. 
 To analyze the BRST cohomology different gradings are
 introduced\,\cite{HenneauxBook,HenneauxReport}: the antifield number
 ($antif$), 
 the pureghost number ($puregh$) and the usual (total) ghost number ($gh$). 
 They are given by, 
$antif(\Phi^i)=0$, $antif(\cB^{*\mu}_A)=1$, $antif(\cC^*_A)=2$,
$puregh(\cB^A_\m)=0$, $puregh(\cC^A)=1$, $puregh(\Phi^*_i) = 0$.
The ghost number is $gh=puregh-antif$ and clearly $gh(S)=0$.  
The assignment of an $antif$ to each variable
is an important feature of the
formalism\,\cite{HenneauxLocality,HenneauxGF1}
that will play a central role later.
For a given $S$, the BRST differential $s$ can be expanded according to the
antifield number\,\cite{Fisch,HenneauxLocality,HenneauxBook,HenneauxGT}. 
In the case of YM \eqref{master} we have
\begin{equation}
\label{eq:delta-gamma} 
s = \delta + \gamma\,,
\end{equation}
where $\delta$ decreases $antif$ by one and $\g$ leaves
it unchanged.
$\d$ is known as the Koszul-Tate differential and  is related to the
EOM of  $S_0$\,\cite{Fisch,HenneauxBook,HenneauxGT},
while $\gamma$ measures the variation of
functions or functionals  along the gauge orbits and it
reduces to the usual off-shell BRS transformation when it
acts in the space of fields $\Phi^i$. 
For $s$ to be nilpotent, we must have $\d^2=\g^2=\d\g+\g\d=0$.
The differential $\delta$ acts only non-trivially on the antifields. 
Due to the choice of $antif$ it gives the
EOM when it acts on $\cB^{*\mu}_A$, i.e.,
$\d \cB^{*\mu}_A = \d S_0/\d \cB_{\mu}^A$, while when it acts
on $\cC^*_A$ it gives $\d\cC^*_A=\cD_\mu^{AB}\cB_B^{*\mu}$ which
ensure the acyclicity of $\d$, i.e., $H_{n>0}(\d)\equiv0$. 
This property is central to the computation of
the BRST cohomology\,\cite{HenneauxLocality,HenneauxGT,HenneauxReport}.
$\d$ implements the EOM in its
cohomology in the following sense. A functional $F$ vanishes
when the EOM of $S_0$ hold if and only if it can be written as a 
$\d$-exact term $F=\d G$, for some functional $G$. Then
$F$ is said to vanish  on-shell (w.\,r.\,t. $S_0$) and  we denote
it by  $F\approx 0$.
Changing the $antif$ of the variables may alter the acyclic property
of $\d$ and the EOM\,\cite{HenneauxGF1,Brandt} that are implemented
in the cohomology and therefore the observables of the theory.

In the BV formalism, a change of gauge
corresponds to a canonical transformation\,\cite{Batalin,Gomis}
using the freedom to add an exact term $(S,\Psi)=\int\!d^n x\,s\Psi$
to the solution of the master equation. Then, $S\rightarrow S^\Psi$ with
\begin{equation}
\label{newbase}
\textstyle S^\Psi=S - (S,\Psi) = S [\,{\Phi^i = 
 \Phi'^i,\Phi^*_i = \Phi'^*_i+\el{\Psi\,}{\Phi^i}}\,],
\end{equation}
where $\Psi$ is the gauge-fixing fermion which must have
$gh(\Psi)=-1$ as the antibracket increases $gh$ by one.
The antifield-independent part of $S^\Psi$ is the ``gauge-fixed''
action $S^\Psi_0$. The $\Psi$'s of interest are those that
completely fix the gauge freedom.
To construct $\Psi$ it is necessary to extend the phase space to
include non-minimal variables $\{\Cb{}^A,b^A\}$ and their respective
antifields.
The $\Cb{}$'s have the
necessary negative $gh$ and are the familiar
antighosts of gauge-fixed actions and should not be confused
with $\cC{}^*$, the antifield of the ghost.
These variables are trivial in the BRST cohomology, i.e., 
$s\Cb{}^A=b^A$, $sb^A=0$, $sb^*_A=-\Cb{}^*_A$ and $s\Cb{}^*_A=0$.
To implement these transformations a new term
$-\Cb{}^*_A b^A$ is added to $S$ in \eqref{master}.

Let $s_\Psi$ be the BRST
transformation generated by $S^\Psi$, $s_\Psi A = (S^\Psi,A)$.
Since $S^\Psi$ is obtained from $S$ by a canonical transformation
the cohomology of $s$ and $s_\Psi$ are isomorphic.
However, \textit{this isomorphism is only guaranteed in the space of
local functionals as long as all antifields are kept}. For YM,
\begin{equation}
\label{gauge-fixed-action1}
S^\Psi =S^\Psi_0 + \textstyle \int
d^n x\,(\cB'^{*\mu}_A\ \gamma_\Psi \cB^{A\mu}
+  \cC'^*_A\ \gamma_\Psi \cC^A - \Cb{}^{\prime*}_{A} b^A)
\end{equation}
where 
$S^\Psi_0=S_0+\!\int\!d^n x\,(\el{\Psi}{\cB^A_\mu}\ \gamma_\Psi \cB^{A\mu}
+ \el{\Psi}{\cC^A}\ \gamma_\Psi \cC^A -
\frac{\d\Psi}{\d\Cb{}^A}b^A)$
is a function of $\cB_\mu^A,\cC^A,\Cb{}^A$ and $b^A$.
The more familiar on-shell BRS transformation in a fixed gauge
specified by $\Psi$ is defined in this formalism by
$s_{\mathrm{BRS}}(\Phi^i) = s_\Psi(\Phi^i)\vert_{\Phi^*_i=0}$.
It corresponds to a global symmetry of the gauge-fixed action
$S^\Psi_0$ and therefore does not require the full
antifield formalism and should not be confused with $s_\Psi$.

 Finally, we discuss the determination of the set of integrated
 observables, i.e., on-shell  gauge invariant
 functionals\,\cite{HenneauxRenorma,HenneauxGT2,HenneauxBook}.  
 Let $A=A(\Phi^i,\Phi^*_i)$ be a local 
 functional with $gh(A)=0$. If $A$ is BRST closed, $sA=0$,  
 by expanding it
 according to $antif$, $A=A_0+\sum_{k\geq 1}A_k$, 
 with $antif(A_k)=k$, we have $\g A_0+\d A_1=0$. 
 Hence, $\g A_0$  vanishes on-shell ($\g A_0\approx 0$ w.\,r.\,t.
 $S_0$)\,\cite{HenneauxGT}. 
 As for the converse, if $\g A_0\approx 0$ w.\,r.\,t. $S_0$
 then  there  exists a BRST-closed functional
 $A$, $sA=0$, with an antifield independent part that equals $A_0$.  
 In this case, $A$ is called a {\em BRST-closed extension of}
 $A_0$. This is a very useful property\,\cite{HenneauxGT2} as it means
 that a given BRST cocycle $A$ is completely determined by its
 antifield independent part $A_0$. 
 Therefore, the set of local gauge invariant functionals can be determined
 by  the cohomology of the  off-shell BRS operator $\g$  modulo
 the EOM for the {\em gauge invariant action} $S_0$
 ($sA=0\iff \g A_0\approx 0$). 

One would wish that similarly, the cohomology of the on-shell BRS
differential can be used  directly to identify the on-shell gauge invariant
operators. However, as first noticed by
Henneaux\,\cite{HenneauxGF1}, this is not the case 
{\em in the space of local functionals} as it requires 
extra conditions\,\cite{HenneauxGF2}. 
Indeed, in a given gauge $\Psi$, there is no guarantee
that a generic local on-shell BRS-closed  functional $A_0$ 
($s_{\mathrm{BRS}} A_0\approx 0$ w.\,r.\,t. $S^\Psi_0$)  
with $gh(A)=0$ possesses a BRST-closed extension $A$. 
Hence, we have to check explicitly for each $A_0$
whether it is possible to find the terms $A_{k\ge1}$, that make
its extension BRST closed, $s_\Psi A=0$.

Whenever we have a BRST extension of an on-shell BRS 
closed local functional $A_0(\Phi)$ in a given gauge, 
then in \textit{any} other gauge $\Psi$ there is an on-shell
BRS-closed functional which corresponds to $A_0$ and can be written as
$A_\Psi(\Phi)=A(\Phi,\Phi^*=\el{\Psi}{\Phi})$.
This illustrates the advantages of the antifield formalism as it enables
us to study the observables of a gauge theory in a manifestly
gauge invariant way. Furthermore,  $\Phi^*=\el{\Psi}{\Phi}$ provides a 
``dictionary'' to interpret the results in any given gauge $\Psi$.
 
Next, we focus on an explicit study of $A_0$ given by 
\eqref{operator}. Using Stokes' theorem,
local functionals can be identified with their integrands modulo total
divergences\,\cite{HenneauxReport}.
Hereafter, we denote the integrand of $A_0$ by $a_0$.
Using the BRST symmetry, we show that $A_0$ is not gauge invariant and
cannot even be extended to an on-shell gauge invariant local functional
(w.\,r.\,t. $S_0$). 
Now, $A_0=\!\!\int\!d^nx\,a_0$ is BRST closed if and only if it can be 
extended to $A=\!\!\int\!d^n x\,a$  in such a way that  it satisfies
the condition  
\begin{equation}
\label{close}
s a +dm = 0,
\end{equation}
where   $a=\sum_{k\geq 0}a_k $  with $antif(a_k)=k$ and $dm$ is the
exterior derivative of some form $m$. However,
$a_0$ cannot be extended to satisfy \eqref{close}. This would
have implied that $\g a_0 \approx 0$, but it follows from 
the integrand of \eqref{operator} that
\begin{equation}
\label{eq:eq-1}
 \g  a_0= \pa^\m( \cC^A \cB^A_\m) + \D a_0,
\end{equation}
where $\D a_0 = -\cC^A (\pa^\m \cB^A_\m - \a f^{ABC}
\cC^B \Cb{}^C +\a b^A)$.
Clearly, this last term does not vanish modulo the EOM
of the gluons and it is not a total derivative because its
Euler-Lagrange derivatives do not vanish (see Theorem 4.1
of\,\cite{HenneauxReport}).
Hence, $A_0$ does not have a BRST-closed extension because 
of the ``obstruction term'' $\D a_0$ of antifield number zero.
Next, we look at the possibility of
deforming the theory in order that $\D a_0$ vanishes modulo the modified
EOM. This can be achieved if in the new action, the EOM for
the auxiliary field $b^A$ are taken to be
\begin{equation}
\label{eq:gauge-condition}
\pa^\m \cB^A_\m-\a f^{ABC} \cC^B\Cb{}^C +\a b^A=0\,.
\end{equation}
These equations constrain the fields $\cB_\m^A$ and completely fix
their gauge freedom. Therefore, the deformation we are seeking 
is a gauge fixing procedure.
In fact, the EOM \eqref{eq:gauge-condition} can only be obtained if we
choose
\begin{equation}
\label{eq:psi-gauge}
\textstyle\Psi = \Cb{}^A (\pa^\m \cB^A_\m - \frac12\a f^{ABC} \cC^B
\Cb{}^C + \frac12\a b^A )\,.
\end{equation}
This is the gauge-fixing fermion that corresponds to the CF gauge.
Hence, it is only in this gauge that $A_0$ is on-shell BRS closed
with respect to the gauge-fixed action.
The restriction to work in a specific gauge is a
natural consequence of the gauge dependence of 
$A_0$. This has been a serious cause of confusion in the
literature on the gauge invariant status of $A_0$.

Some care has to be taken on how to interpret this
deformation\,\cite{Brandt}. This
is not simply a canonical transformation in the
sense of \eqref{newbase} where the theory remains YM.
As we need to use explicitly the EOM of the auxiliary fields, the new
deformed theory is to be seen as a theory where ghosts and
antighosts have now their own dynamics, which is governed by the
action $S_0^\Psi$ that now plays the role of a ``classical action''
\eqref{gauge-fixed-action1}. 
The EOM of $\{\cC,\Cb,b\}$ can
be implemented in the cohomology with a new choice of antifield
number, say $antif_\Psi$, such that 
$antif_\Psi\cB^{*\mu}_A\!=\!antif_\Psi \cC^*_A\!=\!antif_\Psi
\Cb{}^*_A\!=\!antif_\Psi b^*_A\!=\!1$ \cite{HenneauxGF1}. 
We re-emphasize that it is the BRST operator $s$ together with the
original $antif$ grading that is relevant for
questions about gauge invariance and renormalization of local 
operators\,\cite{HenneauxReport}. 

In order to have a better understanding about what prevents $A_0$ to
be gauge invariant, we now try to construct its BRST-closed 
extension\,\cite{HenneauxGF1} by
assuming that the extension exists, $s_\Psi A = 0$.
In line with \eqref{close}, we have for the integrand
$s_\Psi a + dm = 0$.
This equation can be decomposed into a system by inserting the
expansions in $antif_\Psi$, $s_\Psi = \d_\Psi + \g_\Psi$,
$a=\sum_{k\ge0}a_k$ and $m = \sum_{k\ge0}m_k$, as
the resulting terms at each order in $antif_\Psi$ must
vanish separately providing a method of solving it iteratively.

The first two lowest order non-trivial equations are
\begin{eqnarray}
\g_\Psi a_0+\d_\Psi a_1 + dm_0\!\!\!&=&\!\!0\,,\label{eq3}\\
\g_\Psi a_1 +\d_\Psi a_2 + dm_1\!\!\!&=&\!\!0\,.\label{eq4}
\end{eqnarray}
From \eqref{eq4} and by using the nilpotency of $\d_\Psi$ and $\d_\Psi
d+d\d_\Psi=0$ we see that $\g_\Psi a_1$ is $\d_\Psi$-closed modulo
$d$, i.e., $\d_\Psi (\g_\Psi a_1) = d(\d_\Psi m_1)$.
Moreover, by re-expressing (\ref{eq4}) as $\g_\Psi a_1 =
\d_\Psi(-a_2)+d(-m_0)$, $\g_\Psi a_1$ is indeed $\d_\Psi$-exact modulo $d$.
Therefore, $\g_\Psi a_1$ must be a trivial element of the
homology of $\d_\Psi$ in the space of local functionals.

With this necessary condition for $a_1$ in mind we now
try to extend $A_0$ given by \eqref{operator}.
From \eqref{eq:eq-1} and \eqref{eq3} it follows that
$\d_\Psi a_1 = \cC^A(\pa^\m \cB^A_\m - \a f^{ABC} \cC^B
\Cb{}^C +\a b^A)$.

As $\d_\Psi$ acts only on the antifields,
$\d_\Psi a_1$ must be a linear combination of the gauge-fixed EOM.
In this case the only possibility is to invoke the EOM
of $b^A$,
$\d_\Psi b'^*_A = \frac{\d S^\Psi_0}{\d b^A}$, given by
\eqref{eq:gauge-condition}, where $b'^*_a$ are the antifields in the
base $\Psi$ as in \eqref{newbase}. 
This is achieved with the choice of $antif_\Psi$ mentioned above.
Therefore we have $a_1=-\cC^A b'^*_A$ and
\begin{equation}
  \label{eq:eq8}
  -\g_\Psi a_1 = \textstyle{\frac12}\, 
  b'^*_A f^{ABC}\cC^B \cC^C + \Cb^{\prime*}_A \cC^A.
\end{equation}
The r.\,h.\,s. of \eqref{eq:eq8} does not involve any
derivatives of the fields $\{\cA_\m, \cC ,\Cb \}$ so it can not be
$\d_\Psi$-exact as required for the extension to exist.
Thus, we conclude that \eqref{eq:eq8} constitutes an
obstruction for $A_0$ to be extended into a local BRST-closed
functional of $s_\Psi$. In fact, $\g_\Psi a_1$ in
\eqref{eq:eq8} is a non-trivial element of the homology of $\d_\Psi$
with $antif_\Psi\!\!=\!1$ and therefore $\d_\Psi$ mod $d$ is no longer
acyclic.

There is a global symmetry of the gauge-fixed action 
associated to the obstruction \eqref{eq:eq8}.
From the r.\,h.\,s. of \eqref{eq:eq8}, we easily identify the
generators of the symmetry as they couple linearly to the
antifields. Indeed,  any linear function of the antifields is
naturally viewed as a tangent vector to field space
\cite{Witten,HenneauxGT}. 
 The generator of this global symmetry is 
\begin{equation}
\label{s-obs}
\textstyle\frac12 f^{ABC}\cC^B \cC^C\elL{}{b^A}+ \cC^A\elL{}{\Cb{}^A}\ .
\end{equation}
In the BV formalism a  more familiar way to arrive at this symmetry
is to express it as canonically generated in the antibracket, i.e.,
${\hat{\delta}}_\tau(\Phi^i)=(\tau,\Phi^i)$  with  $\tau$ given by
\eqref{eq:eq8}.
 $\hat{\delta}_\tau$ is  one of the two non-diagonal generators of the
 $SL(2,\mR)$ 
symmetry of the gauge-fixed action \eqref{gauge-fixed-action1}.
The functional $A_0$ is also on-shell invariant for the anti-BRS
transformation $\bar s_\Psi$
in the CF gauge. If we attempt to extend $A_0$ into the cohomology of
$\bar s_\Psi$\,\cite{HenneauxGeometric}
we also find an obstruction, in this case it corresponds
to the infinitesimal symmetry 
\begin{equation}
\label{sbar-obs}
\textstyle \frac12 f^{ABC}\Cb{}^B \Cb{}^C\elL{}{b^A}+ 
\Cb{}^A \elL{}{\cC{}^A}\ ,
\end{equation}
which is  the other non-diagonal  generator
${\hat{\delta}}_{\bar{\tau}}$ of the $SL(2,\mR)$ 
symmetry, with $\bar{\tau}=\frac12\, b'^*_A f^{ABC}\Cb{}^B \Cb{}^C
+ \cC^{\prime*}_A \Cb{}^A$.  The third generator,
$\hat{\delta}_{FP}$, is diagonal and is given
by the commutator 
$[{\hat{\delta}}_{{\tau}},{\hat{\delta}}_{\bar{\tau}}]$
of the two non-diagonal generators. $\hat{\delta}_{FP}$ is
the  generator of the ghost number which is
trivially a global symmetry of the gauge-fixed action as $gh(S^\Psi_0)=0$. 

Even if the auxiliary fields are
replaced by their EOM, the obstructions are still present. They will only
involve the ghosts and the antighosts ( $\hat{\delta}_{\tau}\to
\cC^A\elL{}{\Cb{}^A}$, $\hat{\delta}_{\bar\tau}\to\Cb{}^A
\elL{}{\cC{}^A}$ and $\hat{\delta}_{FP}\to \hat{\delta}_{FP}$),
but their algebra will still be $SL(2,\mR)$. The analysis of that case
is found in the work of Brandt \cite{Brandt} where the
CF mass term \eqref{operator} has been studied from the
perspective of the deformations within the extended BRST 
formalism which implements not only
the gauge but also the global symmetries of a given action
\cite{Barnich:2000me}. In this context, the introduction of the CF 
mass term leads to the loss of nilpotency of the
on-shell (anti-)BRS operator due to the modification of the EOM.
More explicitly, $s_{\mathrm{BRS}}^2=\hat\d_\tau\neq0$ and
$\bar s_{\mathrm{BRS}}^2=\hat\d_{\bar\tau}\neq0$. 
This is also linked to the lost of unitarity of the resulting theory
\cite{Brandt}.

To summarize, $A_0$ is on-shell BRS and anti-BRS invariant only in the
CF gauge, but 
if we try to extend this property to any other gauge we encounter
obstructions that are global symmetries of the gauge fixed action.
More precisely, the obstructions \eqref{s-obs} and \eqref{sbar-obs}
to the local extension $A$ of $A_0$ to be, respectively, BRST and
anti-BRST closed, are two of the three generators of $SL(2,\mR)$. 
These  obstructions  are not symmetries of YM as they do 
not involve gauge fields of $S_0$ and their existence is only associated
to the specific choice of the gauge-fixing where $A_0$ is on-shell BRS
closed, namely the CF gauge. In fact, the $SL(2,\mR)$ generators $\hat\d_\tau$
and $\hat\d_{\bar\tau}$ (\ref{s-obs},\ref{sbar-obs})
that constitute the obstruction always involve
variables of the non-minimal sector of the phase space. Hence, these
symmetries are trivial for YM\,\cite{HenneauxReport,HenneauxGT} 
and therefore can not be of any physical relevance.
As $SL(2,\mR)$ is not a symmetry of YM there is
no physical justification to impose to the quantum theory
Ward identities associated with it\,\cite{Lemes:2002jv}.
The expectation that there is a physical meaning attached to the
symmetry breaking of $SL(2,\mR)$ by ghost condensates 
loses all its support in view of this analysis.
We also note that as $A_0$ is not on-shell (BRST) gauge invariant it can
not be used as a mass term to be added to the action.

The $SL(2,\mR)$ symmetry was originally discovered in the gauge-fixed
action for the MA gauge\,\cite{Schaden:1999ew}. 
A similar analysis to the one presented here, can be done by
considering the operator 
$\tilde A_0=\int(\frac12 \cB_{\m }^a \cB^{a\m} +\a \cC^a{\Cb}{}^a)$,
i.e., $A_0$ restricted to the contribution from the off-diagonal fields.
We have also checked that if we set $\a=0$ in \eqref{operator}, 
Landau gauge, we encounter the same obstruction.

Finally, we point out that
the on-shell BRS invariance of $A_0$ does not correspond to a residual
$U(1)^{N-1}$ symmetry due to a partial gauge fixing as claimed in
\cite{Gripaios:2003xq} as there is no gauge freedom left in the CF
gauge. The broader implication of our result is that mass generation in
YM can not be linked to the condensation of the local
functional $A_0$ of dimension two.
\begin{center}
  ACKNOWLEDGMENTS
\end{center}
We thank A. Ach\'ucarro and P. van Baal for discussions
and encouragement. ME also thanks G. Barnich for discussions. The 
research of FF is supported by FOM.

\end{document}